\documentclass[amsmath,amssymb,prl,final,showpacs,twocolumn]{revtex4-1}

\usepackage{bm,hyphenat,xspace}
\usepackage{graphicx,epsfig}
\usepackage{color}

\newcommand{\A}[2]{{}^{#1}\mathrm{#2}}

\begin{document}

\title {Comment on
"Sensitivity of ($d,p$) Reactions to High 
$n$-$p$ Momenta and the Consequences for Nuclear Spectroscopy Studies"
}

\author{A.~Deltuva}
\email{arnoldas.deltuva@tfai.vu.lt}
\affiliation
{Institute of Theoretical Physics and Astronomy,
Vilnius University, Saul\.etekio al. 3, LT-10257 Vilnius, Lithuania
}

\date{3 May, 2018}

 \maketitle

Using  adiabatic distorted-wave approximation (ADWA)
for three-body $(d,p)$ transfer reactions Ref.~\cite{PhysRevLett.117.162502}
claimed that nonlocal optical potentials (NLOP) 
strongly enhance the  sensitivity of low-energy
 cross sections to high-momentum components in the $np$ interaction
and deuteron wave function. This was seen as a 
strong dependence of $(d,p)$  cross sections
on the  $np$ potential model.
In this  comment I demonstrate that a rigorous treatment of three-body 
reactions with NLOP is in sharp contrast with findings of 
 Ref.~\cite{PhysRevLett.117.162502}, but in agreement with
the generally expected low-energy observable insensitivity to short-range
physics.
I use Faddeev theory \cite{faddeev:60a} that, 
though much more complicated than ADWA
in terms of practical implementation \cite{deltuva:09b}, has an advantage 
of providing 
an exact solution of the three-body scattering problem.

\begin{figure}[!]
\begin{center}
\includegraphics[scale=0.66]{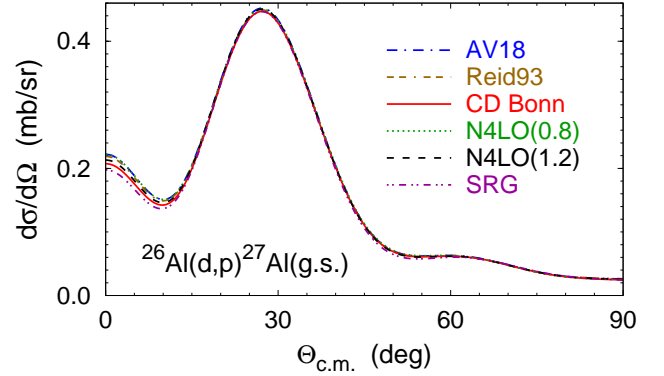}
\end{center}
\caption{\label{fig1}
Differential cross section for the
$\A{26}{Al}(d,p)\A{27}{Al}$(g.s.) reaction
at $E_d=12$ MeV as function of the c.m. scattering angle.
Predictions for different $np$ potentials are compared.}
\end{figure}

I consider the $\A{26}{Al}(d,p)\A{27}{Al}$(g.s.) reaction
at deuteron beam energy $E_d=12$ MeV
where Ref.~\cite{PhysRevLett.117.162502} found the largest
sensitivity to the $np$ potential.
I use same interaction models: $p$-$\A{26}{Al}$ and $n$-$\A{26}{Al}$
NLOP  from Ref.~\cite{giannini} without spin-orbit part and
$\A{27}{Al}$ binding potential with parameters as in 
Ref.~\cite{PhysRevLett.117.162502}.
Beside CD Bonn, Argonne V18 (AV18) and chiral N4LO (with regulators
of 0.8 and 1.2 fm) $np$ potentials
employed in Ref.~\cite{PhysRevLett.117.162502}
I take also Reid93 and 
similarity renormalization group (SRG) potentials. 
The latter broadens the range
of the deuteron $D$-state probability that varies from
2.53\% (SRG) to 5.76\% (AV18).
Results for the differential transfer cross section
are shown in Fig.~\ref{fig1}, they are well-converged
with respect to partial-wave expansion, momentum integration
and Coulomb treatment \cite{deltuva:09b}.
ADWA in Ref.~\cite{PhysRevLett.117.162502} predicted 
a spread of forward-angle cross section by a factor of 2.5 
when using different realistic $np$ potentials but
 in Faddeev-type results the sensitivity is
only 10\% while the magnitude is lower by a factor of
2 to 5 as compared to ADWA. 
For the transfer to
the $\A{27}{Al}$ first excited state $\frac92^+$ (not shown)
the sensitivity is  1\%  and the agreement with ADWA is within 30\%.

Since the accuracy of ADWA with local potentials 
at low-energies was found to be 20\% or better \cite{nunes:11b}
and no sensitivity to the $np$ model was observed
\cite{PhysRevC.95.024603}, 
much larger disagreement in the case of NLOP indicates inadequacy
of ADWA for treating NLOP. The leading-order NLOP effect
can be simulated \cite{PhysRevLett.110.112501}
replacing energy-independent NLOP  by 
the equivalent local optical potential (ELOP)
taken at energy $E_{\rm loc} = E_d/2 + \Delta E$ with 
$\Delta E \approx 40 $ to 70 MeV,
depending on the $np$ potential. Since this represents
a good approximation to the full ADWA  \cite{PhysRevC.95.024603}, 
it makes sense confronting it to the
Faddeev scattering framework \cite{faddeev:60a} to reveal essential
shortcomings that should apply also to the full ADWA for NLOP. 
At a given three-body energy
$E$ in the center-of-mass frame, with $E = E_d\, A/(A+2) - 2.2245$ MeV
for $d+A$ reactions, in the Faddeev formalism
the energy of the interacting nucleon-nucleus subsystem 
$E_{NA} = E - q^2 (A+2)/[2(A+1)m_N]$
is not fixed but depends on the spectator momentum $q$ that
is integration variable
(the same relation holds if one starts from the $(A+2)$ nucleon
problem and considers the effective interaction
between a nucleon and a bound cluster of $A$ nucleons).
However,  ELOP is constructed to be on-shell
equivalent to NLOP at $E_{NA} = E_{\rm loc}$ but deviates
from it at other energies, thus, the two potentials can not
provide an equivalent description of the initial
$d+A$ three-body scattering state. While ELOP may mimic
some features of NLOP, it misses a very important one,
namely, the ability of the energy-independent NLOP to describe the
nucleon-nucleus subsystem over a broader energy range as compared
to ELOP with fixed parameters \cite{giannini}. 
Furthermore, $E_{\rm loc}$ of the order 50 MeV in 
 Ref.~\cite{PhysRevLett.117.162502} at $E_d = 12$ MeV
appears to be very unnatural
from the Faddeev formalism point of view that allows only
 $E_{NA} < 9 $ MeV.

In summary, ADWA  
misses some important features of NLOP. 
Rigorous three-body calculation using NLOP demonstrates
weak dependence of $(d,p)$ cross sections
on the $np$ potential model, in agreement with general expectations
and in sharp contrast with
 Ref.~\cite{PhysRevLett.117.162502}.

\begin{acknowledgments}
A.D. acknowledges the support  by the Alexander von Humboldt Foundation
under Grant No. LTU-1185721-HFST-E.
\end{acknowledgments}


\end{document}